\documentclass{IEEEtran4PSCC}

\ifCLASSINFOpdf
   \usepackage[pdftex]{graphicx}

\else
   \usepackage[dvips]{graphicx}
\fi
\usepackage[cmex10]{amsmath}
\usepackage{amsmath}
\usepackage{amssymb,,amsthm}

\newtheorem{lemma}{Lemma}
\theoremstyle{definition}

\usepackage{algorithmic}
\usepackage{algorithm}

\usepackage{booktabs}
\usepackage{multirow}
\usepackage{xcolor}

\DeclareMathOperator*{\argmin}{arg\,min}

\makeatletter
\let\old@ps@headings\ps@headings
\let\old@ps@IEEEtitlepagestyle\ps@IEEEtitlepagestyle
\def\psccfooter#1{%
    \def\ps@headings{%
        \old@ps@headings%
        \def\@oddfoot{\strut\hfill#1\hfill\strut}%
        \def\@evenfoot{\strut\hfill#1\hfill\strut}%
    }%
    \def\ps@IEEEtitlepagestyle{%
        \old@ps@IEEEtitlepagestyle%
        \def\@oddfoot{\strut\hfill#1\hfill\strut}%
        \def\@evenfoot{\strut\hfill#1\hfill\strut}%
    }%
    \ps@headings%
}
\makeatother

\psccfooter{%
        \parbox{\textwidth}{\hrulefill \\ \small{24th Power Systems Computation Conference} \hfill \begin{minipage}{0.2\textwidth}\centering \vspace*{4pt} \includegraphics[scale=0.06]{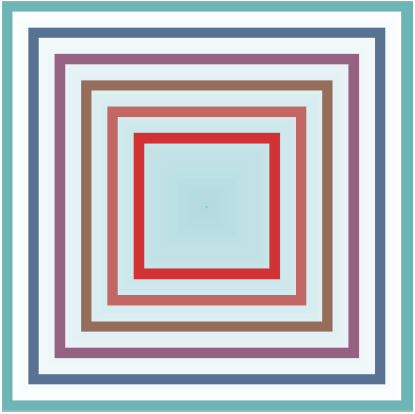}\\\small{PSCC 2026} \end{minipage} \hfill \small{Limassol, Cyprus --- June 8-12, 2026}}%
}

\usepackage[backend=biber,
style=ieee,
citestyle=numeric-comp,
sorting=none,
doi=false,isbn=false,url=false ]{biblatex}

\IEEEoverridecommandlockouts
\begin{document}

\title{Optimal Kron-based Reduction of Networks (Opti-KRON) for Three-phase Distribution Feeders}

\author{
\IEEEauthorblockN{Omid Mokhtari, Samuel Chevalier, Mads Almassalkhi}
\IEEEauthorblockA{Department of Electrical and Biomedical Engineering \\
University of Vermont \\
Burlington, VT, USA \\
\{omid.mokhtari, schevali, malmassa\}@uvm.edu}

\thanks{This material is based upon work supported by the U.S. Department of Energy’s Office of Energy Efficiency and Renewable Energy (EERE) under the Enabling Place-Based Renewable Power Generation using Community Energyshed Design initiative, award number DE-EE0010407. The views expressed herein do not necessarily represent the views of the U.S. Department of Energy or the United States Government.}}

\maketitle

\begin{abstract}
This paper presents a novel structure-preserving, Kron-based reduction framework for unbalanced distribution feeders. The method aggregates electrically similar nodes within a mixed-integer optimization (MIP) problem to engender reduced networks that optimally reproduce the voltage profiles of the original full network. To overcome computational bottlenecks of MIP formulations, we motivate an exhaustive-search formulation to identify optimal aggregation decisions, while enforcing voltage margin limits. The proposed exhaustive network reduction algorithm is parallelizable on GPUs, which enables scalable network reduction. The resulting reduced networks approximate the full system’s voltage profiles with low errors, and are suitable for steady-state analysis and optimal power flow studies. The framework is validated on two real utility distribution feeders with $5{,}991$ and $8{,}381$ nodes. The reduced models achieve up to $90\%$ and $80\%$ network reduction, respectively, while the maximum voltage-magnitude error remains below $0.003$ p.u. Furthermore, on a 1000-node version of the network, the GPU-accelerated reduction algorithm runs up to 15x faster than its CPU-based counterpart.
\end{abstract}



\section{Introduction}
The rapid integration of distributed energy resources (DERs) and advanced control schemes has significantly increased the complexity of distribution networks, posing challenges for optimization and real-time decision-making~\cite{DOE_report}. Detailed network models with thousands of nodes are computationally intensive and are impractical for large-scale optimal power flow and control problems. Consequently, research on network reduction has emerged, where proposed network reduction tools seek to simplify network models while preserving essential electrical characteristics.

One network reduction technique is bus elimination via Kron reduction~\cite{kron_dorfler}, where selected nodes are eliminated via the Schur complement of the nodal admittance matrix, resulting in an equivalent network that preserves the impedance seen between kept nodes. However, the set of reducible nodes is restricted to those with zero current injection.
References~\cite{Adaptive_Network_Science, dnr_ISGT_1ph} develop network reduction methods for single-phase distribution systems. In~\cite{Adaptive_Network_Science}, the network parameters of a reduced model are updated through optimization, whereas~\cite{dnr_ISGT_1ph} reduces the feeder by recursively merging parts of the radial graph into equivalent nodes. However, both approaches are formulated for single-phase systems and therefore do not extend to real-world unbalanced three-phase networks.

For unbalanced three-phase distribution feeders, several reduction approaches have been proposed. In~\cite{multiphase_reduction, inv_reduction}, reduced feeders are constructed by eliminating non-critical buses from the feeder impedance matrices and reconstructing the corresponding network elements for the kept nodes.
Similarly,~\cite{todorovski2024, desouza2021, segment_substitution_3ph_unb_dd} construct reduced networks by replacing selected parts of the feeder with simplified line-based representations. In~\cite{todorovski2024, segment_substitution_3ph_unb_dd}, the reduced network is represented by a single segment between two remaining nodes, whereas~\cite{desouza2021} replaces multiple sectors, each with one line.
A graph-based reduction approach is presented in~\cite{MST_2014_3ph_unb_he}, where a minimum-spanning-tree search is used to identify feeder sections for aggregation into a smaller reduced network.
Application-specific reduced feeders are developed in~\cite{access_CASOLINO_3ph_unb_sens, real_time_3ph_unb_heu, Kar2024}, where selected buses are retained and the remaining network is reduced for power flow analysis, real-time simulation, or risk-aware islanding studies.
The existing network reduction methods discussed so far assume a predefined set of nodes to be retained in the reduced network. However, the quality of the reduced network depends strongly on that choice. In general, more aggressive network reduction leads to larger approximation errors. Therefore, incorporating an error metric into the reduction process to determine the set of kept nodes and balance reduction level against accuracy is important. However, this aspect is lacking in existing network reduction methods for three-phase feeders.


In prior works~\cite{optikron2025} and~\cite{optiKRON2022}, an optimal Kron-based reduction of networks (Opti-KRON) framework has been developed for single-phase balanced feeders, where the reduced networks do not rely on any predefined set of nodes and explicitly trade off reduction level against accuracy via tunable bounds on voltage magnitude error. A mixed-integer linear program (MILP) was proposed to identify the optimal set of nodes and clusters that maximize reduction while minimizing voltage deviations across representative loading scenarios. In addition, because Kron reduction can yield dense, meshed networks~\cite{kron_dorfler},~\cite{optikron2025}, the framework introduced a radialization step to ensure the reduced network remains radial.
However, the previous Opti-KRON formulation is limited to single-phase networks. 
To the best of our knowledge, the literature lacks a scalable approach that computes optimal network reductions for unbalanced three-phase distribution feeders. 

In this paper, we extend the Opti-KRON framework of~\cite{optikron2025} to unbalanced three-phase distribution networks. We develop a three-phase Kron-based reduction and radialization procedure.
To formulate the optimal reduction, we first generalize the single-phase Opti-KRON model to three-phase feeders.
Furthermore, we introduce an exhaustive-search algorithm, with the MILP formulation serving as a methodological baseline.
The proposed exhaustive search offers several advantages. First, it operates directly on complex values without requiring decomposition into rectangular coordinates. Second, it can evaluate nonconvex objective functions. Third, its structure supports parallel computation on GPUs.  
Consequently, the proposed method achieves significant computational speedups on large-scale feeders while increasing reduction accuracy. This paper makes the following contributions:
\begin{itemize}
    \item Three-phase MILP extension: The single-phase optimal Kron-based network reduction is extended to unbalanced three-phase radial feeders.
    
    \item Exhaustive search framework: An alternative exhaustive search based algorithm, with the ability to handle nonconvex objective functions, is proposed to increase scalability and accuracy.

    \item Three-phase radialization: The radialization procedure is extended to the three-phase setting to ensure that the reduced networks preserve both topological structure and phase connectivity.
\end{itemize}

The remainder of this paper is organized as follows. In Section~\ref{sec:model}, the three-phase network model and Kron reduction are summarized. The concept of Kron-based network reduction is introduced in Section~\ref{sec: kron-based}. Section~\ref{sec:search} introduces our exhaustive-search framework for optimal Kron-based reduction.
We then describe the radialization step in Section~\ref{sec:radialization}. Experimental results are presented in Section~\ref{sec:results}. We conclude in Section~\ref{sec:end} with a summary and future directions.

\section{Three-Phase Model and Kron Reduction}\label{sec:model}
\subsection{Network Representation}
Consider a distribution network as a graph $\mathcal{G}=(\mathcal{V},\mathcal{E})$ with $n=\lvert\mathcal{V}\rvert$ nodes and $m=\lvert\mathcal{E}\rvert$ branches. We define the network topology via the adjacency matrix $\Lambda \in \{0,1\}^{n \times n}$, where $\Lambda_{ij}=1$, if a branch connects nodes $i$ and $j$, and $\Lambda_{ij}=0$ otherwise. Each node $i$ carries a subset of phases $\phi_i\subseteq\{a,b,c\}$. 
Each node $i$ is associated with a three-phase voltage vector $V_i\in\mathbb{C}^3$ and current injection vector $I_i\in\mathbb{C}^3$. Complex nodal voltage and current injection vectors are defined as
\[
  V = \begin{bmatrix} V_1  \\ \vdots \\ V_n \end{bmatrix} \in \mathbb{C}^{3n}, 
  \qquad
  I = \begin{bmatrix} I_1 \\ \vdots \\ I_n \end{bmatrix} \in \mathbb{C}^{3n}.
\]
The nodal admittance matrix $Y \in \mathbb{C}^{3n \times 3n}$ has a block structure, where each block $Y_{ij} \in \mathbb{C}^{3 \times 3}$ represents the self and mutual phase couplings between phases of nodes $i$ and $j$. 
Phase absence at node $i$ is represented by zeroing out the rows and columns of $Y$ that correspond to the missing phases in the blocks associated with $i$, and by setting the corresponding entries of $V$ and $I$.~\footnote{Equivalently, one may remove absent-phase rows/columns and work with an $n_\phi\times n_\phi$ matrix indexed by existing node–phase pairs; both views are mathematically equivalent.}
Kirchhoff’s Current Law is enforced via
\begin{align} 
  I \;=\; Y\,V. \label{eq:KCL-3phase}
\end{align}

\subsection{Kron Reduction of Three-Phase Networks}\label{kron-reduction}
Let the node set be partitioned into $\mathcal{K}$, the nodes to be \emph{kept}, and $\mathcal{R}$, the nodes to be \emph{reduced} with $I_i = \mathbf{0}, \ \forall i \in \mathcal{R}$.
We can partition~\eqref{eq:KCL-3phase} into
\begin{align} \label{eq: Y-partitioned}
        \left[\begin{array}{c}
            {I}_{\mathcal{K}}\\
              \mathbf{0}
        \end{array}\right] & =\left[
        \begin{array}{cc}
            Y_{\mathcal{K}\mathcal{K}} & Y_{\mathcal{K}\mathcal{R}} \\
            Y_{\mathcal{R}\mathcal{K}} & Y_{\mathcal{R}\mathcal{R}}
        \end{array}
        \right]\left[\begin{array}{c}
        {V}_{\mathcal{K}}\\
        {V}_{\mathcal{R}}
            \end{array}\right].
\end{align}
The Kron reduction of $Y$ is the Schur complement of $Y_{\mathcal{R}\mathcal{R}}$~\cite{kron_dorfler}. Thus, the Kron reduced admittance matrix, denoted as $Y_{\rm Kron}$, is given by
\begin{align} \label{eq: Y_kron}
     Y_{\rm Kron} \;=\; Y_{\mathcal{K}\mathcal{K}} - Y_{\mathcal{K}\mathcal{R}}\; {Y^{+}_{\mathcal{R}\mathcal{R}}} \; Y_{\mathcal{R}\mathcal{K}},
\end{align}
where $\{ \cdot \}^+$ represents the Moore-Penrose Pseudoinverse~\footnote{The pseudo-inverse is utilized to account for zero rows and columns associated with missing phases during elimination. } and
\begin{align}
  I_\mathcal{K} \;=\; Y_{\mathrm{Kron}} \; V_\mathcal{K}. \label{eq:Ik-YkVk-3phase}
\end{align}
In this work, Kron reduction is applied at the node level for three-phase systems, meaning that when a node is reduced, all of its phases are simultaneously reduced. 
Under this assumption, $Y_{\mathrm{Kron}} \in  \mathbb{C}^{3|\mathcal{K}| \times 3|\mathcal{K}|}$.

\subsection{Clustering for Kron-based Reduction}
\begin{figure}
    \centering
    \includegraphics[width=1\linewidth]{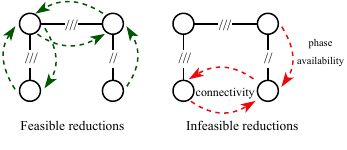}
    \caption{Examples of allowable and unallowable (infeasible) reductions.}
    \label{fig: infeas_red}
\end{figure}

\begin{figure}
    \centering
    \includegraphics[width=1\linewidth]{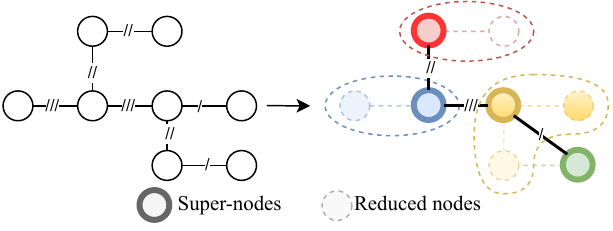}
    \caption{Example of the assignment process, i.e., each reduced node transfers its injection to its assigned super-node.}
    \label{fig:assignment_example}
\end{figure}

As shown in \eqref{eq: Y-partitioned} and \eqref{eq:Ik-YkVk-3phase}, Kron reduction preserves voltage profile when the reduced nodes have zero current injection.
To enable reduction beyond strictly zero-injection nodes, we introduce an \emph{assignment} scheme, where each reduced node moves its current injection to a kept adjacent node. We refer to the kept nodes $\mathcal{K}$ as \emph{super-nodes}; each super-node contains its own current injection together with those of the reduced nodes assigned to it. Each super-node $k \in \mathcal{K}$ represents a \emph{cluster} consisting of itself and its assigned \emph{children-nodes}.
These assignments, however, must obey several rules:  
\begin{enumerate}

\item{Connectivity:} each child must be connected to its super-node through existing branches, so that every cluster forms a connected sub-graph of the original network.

\item{Phase availability:} a reduced node $i$ can only be assigned to a super-node $k$ if $\phi_i \subseteq \phi_k$.

\item{Uniqueness:} every reduced node $i \in \mathcal{R}$ is assigned to exactly one super-node.
\end{enumerate}
Fig.~\ref{fig: infeas_red} visualizes examples of feasible and infeasible reductions. 
Fig.~\ref{fig:assignment_example} illustrates the assignment procedure. This example demonstrates how assignments create disjoint, connected clusters that collectively cover the entire network.
To formalize the assignment of reduced nodes to super-nodes, we introduce a binary assignment matrix $A \in \{0,1\}^{n \times n}$, where $A_{i,j}=1$, if node $j$ is assigned to node $i$. Additionally, $A_{i,i}=1$ if node $i$ is a super-node and  $A_{i,i}=0$ otherwise. Accordingly, the aggregated current vector under an arbitrary assignment matrix $A$ is obtained by applying $A$ in block form to the three-phase injections:
\begin{align} \label{eq:i_agg}
    I^{\rm agg} \;=\; (A \otimes \mathbf{I}_{3})\, I \, \in \mathbb{C}^{3n},
\end{align}
where $\mathbf{I}_{3}$ is the $3\times 3$ identity matrix and $\otimes$ denotes the Kronecker product. Now, the KCL equation can be stated as
\begin{align} \label{eq:v_agg}
    (A \otimes \mathbf{I}_{3})\; I \;=\; Y \; V^{\rm agg}.
\end{align}
Now, the \textit{Kron-based} reduced network is achieved by reducing the set of nodes $\mathcal{R} = \{i \in \mathcal{V}|I^{\rm agg}_i = \mathbf{0}\}$.
The assignment of nodes with nonzero injections generally leads to $V^{\rm agg} \neq V$.
This, in turn, results in an error in voltages between nodes in the reduced and full networks. Consequently, in the next section, we aim to find the optimal assignment matrix to maximize number of reduced nodes and accuracy simultaneously.

\section{Optimal Kron-based Reduction of Networks} \label{sec: kron-based}
 The choice of $A$ determines how injections are aggregated and how voltages are represented, and thus directly affects both the reduction level and the accuracy of the reduced model. In general, greater reduction introduces larger errors.
 The objective is to find the assignment matrix $A$ by finding a desired balance between reduction and accuracy.
 One way to achieve this is to cast the search for $A$ as an optimization problem, where the assignment rules -- phase availability, connectivity, and uniqueness -- are enforced as constraints, as was done in Opti-KRON~\cite{optikron2025} for single phase networks. 
 
 In this section, we extend the formulation to three-phase feeders. 
 The core optimization framework follows the single-phase structure but is reformulated to handle three-phase feeders.
 Let $\mathcal{L}$ denote a library of operating scenarios given from AC load flow solutions on the full network. For each scenario $l \in \mathcal{L}$, $\hat{V}_l \in \mathbb{C}^{3n}$ and $\hat{I}_l\in \mathbb{C}^{3n}$ are the three-phase nodal voltage and current injection vectors of the full network that satisfy $Y \hat{V}_l = \hat{I}_l$. The voltage of the network under aggregated current is denoted by $V_l$, such that $Y V_l = A\hat{I}_l$.
 In this paper, error is defined as the difference between the nodal voltage in the full and reduced networks. In the reduced network, super-node voltages are obtained through~\eqref{eq:v_agg}, while each reduced node is assigned the voltage of its associated super-node. Accordingly, for $l \in \mathcal{L}$, error is computed as
 \begin{align} \label{eq: error}
 {E_{l}} = \overbrace{(A \otimes \mathbf{I}_{3}) {\rm diag}\{\hat{V}_l\}}^{\text{Full}} - \overbrace{{\rm diag}\{V_l \} (A \otimes \mathbf{I}_{3})}^{\text{ Reduced}} \in \mathbb{C}^{3n\times 3n}.
\end{align}
From~\eqref{eq: error}, $A_{i,j}=1 \rightarrow E_{l,i,j} = \hat{V}_{l,j} - V_{l,i}$ and $A_{i,j}=0 \rightarrow E_{l,i,j}=0$. To quantify the maximum voltage deviation within each cluster and evaluate the quality of a given reduction, we define the maximum intra-cluster error (MICE) as
\begin{align}
    \text{MICE}_{i,l} = \left\Vert \vec{e}_i^T E_{l} \right\Vert_{\infty}.
\end{align}
Note that $\vec{e}_i$ represent the $i^{\rm th}$ standard basis vector. Using MICE as the measure of accuracy, we formulate the three-phase Opti-KRON problem to find the assignment matrix as follows.
\begin{subequations} \label{raw model}
        \begin{align}
    \min_{A} \quad &{\sum_{l \in \mathcal{L}} {\sum_{i \in \mathcal{V}} { \left\Vert \vec{e}_i^T E_{l} \right\Vert_{\infty} }}  - \alpha \sum_{i \in \mathcal{V}} (1- A_{i,i}) }  \label{eq: obj2}\\
    {\rm s.t.}\;\; \quad & {E_{l}} = (A \otimes \mathbf{I}_{3}) {\rm diag}\{\hat{V}_l\} - {\rm diag}\{V_l \} (A \otimes \mathbf{I}_{3}) \quad  \label{eq: error-opt}\\
    & \mathbf{Y} V_l = (A \otimes \mathbf{I}_{3}) \hat{I_l}  \label{eq: powerflow}\\
    & - \bar{E} \leq |(A \otimes \mathbf{I}_{3})^{\rm T}V_l|-|\hat{V}_l| \leq \bar{E}  \label{eq: v_abs}\\   
    & A^{\rm T} \mathbf{1} = \mathbf{1} \label{eq: assign_limit1}\\
    &A_{i,j} \leq A_{i,i} \quad \forall i,j \in \mathcal{V} \label{eq: assign_limit2}\\
    &A_{i,j} \leq \Lambda_{i,j} \quad \forall i,j \in \mathcal{V}, i\neq j\label{eq: assign_adj}\\
    &A_{i,j} = 0 \quad \forall i,j \in \mathcal{V}, \phi_j \nsubseteq \phi_i \label{eq: phase availability}\\
    &\sum_{i \in \mathcal{V}}(1-A_{i,i}) \leq  q \label{eq: cutting plane}\\
    &A_{i,j} \in \{0,1\} \quad \forall i,j \in \mathcal{V}. \label{eq :binary}
	\end{align}
 \end{subequations}
The objective function~\eqref{eq: obj2} minimizes the sum of MICE values across all scenarios, while simultaneously minimizing the number of super-nodes. The trade-off is tuned by the parameter $\alpha$. The KCL equation is enforced in~\eqref{eq: powerflow}. 
Voltage magnitude deviations are constrained in~\eqref{eq: v_abs} to lie within a tolerance~$\bar{E}$.
Unique assignments are enforced by~\eqref{eq: assign_limit1} to ensure each reduced node maps to exactly one super-node.
Equation~\eqref{eq: assign_limit2} allows assignments only to super-nodes.
To keep the clusters connected,~\eqref{eq: assign_adj} restricts clustering to adjacent nodes through the adjacency matrix $\Lambda$, while~\eqref{eq: phase availability} enforces availability of the reduced node phases at the super-nodes.
The constraint~\eqref{eq: cutting plane} bounds the number of reduced nodes, to speed-up the solve times.
Finally,~\eqref{eq :binary} declares all assignment variables as binary.

To solve~\eqref{raw model}, we follow the methodology adapted from~\cite{optikron2025}, which addressed the same challenges:
\begin{enumerate}
    \item The problem is a mixed-integer nonlinear program (MINLP), due to the nonlinear voltage constraints~\eqref{eq: error-opt} and~\eqref{eq: v_abs}. This can be relaxed to a mixed-integer linear program (MILP) by introducing auxiliary variables and Big~M method. Accordingly, the resulting MILP should be interpreted as a relaxation of the original nonconvex MINLP: feasibility is imposed through linearized voltage magnitude error constraints, while the exact voltage errors are evaluated a posteriori.
    
    \item All continuous variables are expressed in complex form, which prevents us from solving it with standard solvers. There are two possible formulations to address this. In the first approach, the voltage magnitude is expressed as $|V_i| = ||V^{\rm real}_i + j V^{\rm imag}_i||_2$, which adds a nonconvex and nonlinear constraint. Otherwise, we can  adopt the standard rectangular decomposition to decompose complex variables into real and imaginary components and approximate~\eqref{eq: v_abs}. In this work, we adopt the rectangular formulation to preserve linearity.
    
    \item The reduction level is limited by~\eqref{eq: assign_adj} and~\eqref{eq: cutting plane}. Therefore, we solve the problem iteratively: at each stage, the model is resolved with an updated adjacency matrix that reflects the evolving cluster structure. The process terminates when no reduction occurs, i.e., no further reduction satisfies the voltage error threshold.
\end{enumerate}

To iteratively solve~\eqref{raw model}, the adjacency matrix must be updated after each iteration to reflect newly admissible assignments. After each iteration $t$, based on how reduced nodes are assigned, we update the adjacency matrix to model connections among super-nodes as 
\begin{align}\label{eq: update_adj}
   \Lambda^{(t+1)}= (A^* \Lambda {A^*}^{\rm T}) \odot (\mathbf{1} \mathbf{1}^{\rm T} - I),
\end{align}
where, $A^*$ represents the optimal assignment matrix obtained at iteration $t$, and $\odot$ denotes element-wise multiplication.
~\eqref{eq: update_adj} connects each super-node to the neighbors of itself and its assigned nodes. If the original network is radial, $\Lambda$ remains radial throughout. The number of reduced nodes $n_r^{(t)}$ and super-nodes $n_s^{(q,t)}$ at iteration $t$ are both functions of the reduction limit $q$. 
The number of nodes that can be reduced in iteration $t$ is bounded by $n_r^{(t)} \leq q$,
and the number of remaining super-nodes evolves as
\begin{equation}
    n_s^{(q,t)} = n - \sum_{\tau=1}^{t} n_r^{(\tau)}.
\end{equation}
Since $\Lambda^{t}$ is radial, the number of lines at each iteration is $n_s^{(q,t)} - 1$. Accordingly, the number of \textit{distinct} feasible reduction scenarios restricted by~\eqref{eq: assign_adj} equals $2(n_s^{(q,t)} - 1)$, i.e., two scenarios per available line (for every line, the load could move across in either direction). Here, by distinct feasible reduction scenarios, we mean the total number of pairs $(i,j)$ that satisfy~\eqref{eq: assign_adj}.

\begin{figure}
    \centering
    \includegraphics[width=0.8\linewidth]{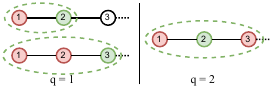}
    \caption{Example of assignment evolution under different reduction limits:
    Assume the best reduction configuration is when nodes 1 and 2 are assigned to node 3. Because of~\eqref{eq: assign_adj}, node 1 cannot be directly reduced to node 3 in a single iteration.
    However, when $q=2$, the algorithm behaves greedily and attempts to reduce two nodes simultaneously. Consequently, it fails to reach the optimal reduction.}
    \label{fig: lemma}
\end{figure}

The number of iterations in~\eqref{raw model} is influenced by the reduction limit $q$ in~\eqref{eq: cutting plane}.
 Setting $q$ to a larger value allows the optimizer to make more reductions per Opti-KRON iteration, leading to potentially fewer total iterations until convergence is achieved. However, since reductions can only be made on direct neighbors at each iteration (due to the connectivity constraint \eqref{eq: assign_adj}), we have observed that the optimizer can start to make ``greedy" decisions for $q>1$. Specifically, it can trade long-term reduction quality (in terms of voltage prediction error) for short-term nodal reduction gains which, while feasible and locally optimal, eliminate the possibility of higher quality reduction decisions on future iterations. This will be illustrated directly in Fig. \ref{fig: lemma}. This behavior arises because larger $q$ values reduce the number of distinct feasible reduction scenarios that can be explored.
When reducing many nodes at one step, $n_s^{(q,t)}$ decreases sharply and leads to a smaller set of distinct feasible choices. The following lemma formalizes the impact of $q$ on the total number of feasible reduction scenarios explored throughout the reduction process.

\begin{lemma} \label{lemma: cp}
Consider iteratively reducing a network of size $n$ by $\Delta n$ nodes using~\eqref{raw model}. Let $T(q)$ denote the number of iterations required to complete the reduction for a given reduction limit $q$. Let $W^{T(q)}$ denote the total number of distinct feasible reduction assignments explored over those $T(q)$ iterations. Then, 
\begin{align}
    W^{T(1)} \geq W^{T(q_{2})},  \quad \  \forall q_2>1.
\end{align}
\end{lemma}
\begin{proof}
Consider that for $q_1 = 1$, $T(q_{1}) = \Delta n$, while for $q_2 > 1$, $T(q_{2}) \le \Delta n$, which results in $T(q_1) \ge T(q_2)$. At each iteration $t$, the number of super-nodes satisfies:
\begin{align}
\overbrace{n - q t}^{\text{up to}~q \text{ nodes per iteration}} \leq n_s^{(q,t)} \leq \overbrace{n - t}^{\text{one node per iteration
}}.
\end{align}
This implies $n_s^{(q_1,t)} \ge n_s^{(q_2,t)}$ for all $t \le T(q_2)$. The number of feasible reductions at each iteration is $2(n_s^{(q,t)} - 1)$, and therefore:
\begin{subequations}
\begin{align}
 W^{T(q_{1})} &:= \sum_{t=1}^{T(q_1)} 2(n_s^{(q_1,t)} - 1) 
\\
 W^{T(q_{2})}  &:= \sum_{t=1}^{T(q_2)} 2(n_s^{(q_2,t)} - 1).
\end{align}
\end{subequations}
Since $n_s^{(q_1,t)} \ge n_s^{(q_2,t)}$ and $T(q_1) \ge T(q_2)$, then $W^{T(q_1)} \geq W^{T(q_2)} $.
\end{proof}
Note that Lemma~\ref{lemma: cp} guarantees $W^{T(q_1)} \geq W^{T(q_{2})}$ only for $q_1=1$ and $q_2>1$ and not generally for $q_1>1$ and $q_2>q_1$.
Lemma~\ref{lemma: cp} confirms that smaller reduction limits, particularly $q=1$, enable exploration of a larger set of feasible reduction scenarios, which leads to more informative and potentially higher-quality network reductions. 
 Additionally, with $q=1$, at iteration $t$ the search tree becomes shallow, i.e., there are $2(n-t)$ reduction scenarios that can be evaluated independently. This motivates a parallel exhaustive search scheme in which each candidate reduction is assessed by solving a linear system of equations. In the next section, we explain the details of the proposed exhaustive search and its parallel implementation. 

\section{Exhaustive Search for Reduction}\label{sec:search}
Building on the MILP formulation of Section III, which optimally determines reduction assignments, we now introduce a complementary approach designed for parallel execution on GPUs. 
Motivated by Lemma~\ref{lemma: cp}, reducing one node per iteration allows us to evaluate each distinct reduction scenario independently and in parallel. 
Additionally, while solving the MILP requires decomposing complex variables into their real and imaginary parts, the exhaustive search approach operates directly in the complex domain and enables exact computation of voltage deviations.
Furthermore, unlike the MILP, which minimizes a linear objective, the exhaustive-search framework allows direct evaluation of nonlinear error measures. In particular, objectives based on voltage magnitudes, while inherently nonconvex, can be assessed exactly for each candidate reduction. Accordingly, we adjust the objective of each iteration to minimize the sum of MICE (SMICE), where error is measured as voltage \textit{magnitude} deviation. At each iteration, we find the assignment such that
\begin{equation}
    (s^{*}, r^{*}) =
    \arg\min_{(s,r)\in \mathcal{A}^{(t)}}
     \big\|\,|\hat{\mathbf{V}}| - |\mathbf{V}^{c}_{(s,r)}|\,\big\|_\infty,
    \label{eq:iter_obj}
\end{equation}
where $\mathbf{V}^{c}_{(s,r)}$ denotes the nodal voltages obtained after assigning node $r$ to super-node $s$ and $\mathcal{A}^{(t)}$ represents the set of all distinct feasible assignments at iteration $t$.
Algorithm~\ref{alg: exhaustive-search} describes this procedure in detail. 

\begin{algorithm}[!t]
	\caption{Exhaustive Network Reduction Algorithm}
    \begin{algorithmic}[1]
	\STATE \textbf{Input}: $Y$ (admittance matrix), $\hat{V}_l $ (voltage data), $\hat{I}_l$ (current injection data), $\Lambda$ (adjacency matrix), $\bar{E}$ (voltage error bound), $\phi$ (set of available phases)
            \STATE  $\Delta \gets \infty$ \hfill $\triangleright$ Termination indicator
            \STATE  $I \gets \hat{I}$ \hfill $\triangleright$ Aggregated current injections
            \STATE  $\mathcal{S} \gets \{(i,i)| i \in \mathcal{V} \}$ \hfill $\triangleright$ Initialize assignment set
            
             \WHILE{$\Delta \ne 0$ } 
                    \STATE $\mathcal{A} = \{ (s,r)|\Lambda_{s,r} = 1 \; \&\; \phi(r) \subseteq \phi(s) \}$ \hfill $\triangleright$ Candidates 

                    \noindent\makebox[\linewidth]{\rule{0.2\linewidth}{0.4pt} \hspace{0.5em} Parallel computation block  \hspace{0.5em} \rule{0.2\linewidth}{0.4pt}}

                    \FORALL{ $(s,r) \in \mathcal{A}$}
                        \STATE $I^c \gets I$ \hfill $\triangleright$ Initial current vector
                        \STATE $I^c_{s} \gets I^c_{s} + I^c_{r}  $  \hfill $\triangleright$ Aggregate current
                        \STATE $I^c_{r} \gets 0  $
                        \STATE solve $I^c  = Y \; V^c$  \hfill $\triangleright$ Find new voltage
                        \STATE $V^c_{r} \gets V^c_{s}$  \hfill $\triangleright$ Assign voltage for new reductions
                        \FOR{$(\mathfrak{s}, \mathfrak{r}) \in \mathcal{S}$}  
                            \STATE $V^c_{\mathfrak{r}} \gets V^c_{\mathfrak{s}}$ \hfill $\triangleright$ Assign previously reduced
                        \ENDFOR
                        
                        \STATE $e^c = ||\hat{V}| - |V^c| |$ \hfill $\triangleright$ Phase to phase error
                        \IF{$\max_{i \in \mathcal{V}} e^c_i \le \bar{E}$}
                            \FOR{$i \in \mathcal{V}$}
                                \STATE $\mathcal{C}_i = \{r \in \mathcal{V} | (i,r) \in ((s,r) \cup \mathcal{S})  \}$  $\triangleright$ Clusters
                                \STATE ${\rm MICE}_i = \max_{r \in \mathcal{C}_i} e^c_r$
                            \ENDFOR
                            \STATE $\text{SMICE}_{s,r} = \sum_{i \in \mathcal{V}} {\rm MICE}_i$
                        \ELSE
                            \STATE $\text{SMICE}_{s,r} = \infty$
                        \ENDIF
                        \ENDFOR 
\vspace{3pt}
{\hrule  }
\vspace{3pt}
                    \IF{$\min_{(s,r) \in \mathcal{A}} \text{SMICE}_{s,r} \neq \infty$}
                    
                    \STATE $(\mathfrak{s}, \mathfrak{r}) = \argmin_{(s,r) \in \mathcal{A}} \text{SMICE}_{s,r}$
                    
                    \STATE $\mathcal{S} \gets \mathcal{S}\setminus(\mathfrak{r},\mathfrak{r})$  \hfill $\triangleright$ Remove the reduced node
                    
                    \STATE $\mathcal{S} \gets \mathcal{S} \cup (\mathfrak{r},\mathfrak{s})$ \hfill $\triangleright$ Add the new assignment
                    
                    \STATE ${C}_\mathfrak{r} \gets \{ (\mathfrak{r},r)|(\mathfrak{r},r) \in \mathcal{S} \} $ \hfill $\triangleright$ old cluster of $\mathfrak{r}$
                    
                    \STATE ${C}_\mathfrak{s} \gets \{ (\mathfrak{s},r)|(\mathfrak{r},r) \in \mathcal{S} \} $ \hfill $\triangleright$ new cluster

                    \STATE $\mathcal{S} \gets (\mathcal{S} \setminus {C}_\mathfrak{r}) \cup {C}_\mathfrak{s}$ \hfill $\triangleright$ merge clusters

                    \STATE $A \gets \mathbf{0}_{n \times n} $
                    
                    \FORALL{$(s,r) \in \mathcal{S}$}
                        \STATE $A_{s,r} = 1 $ \hfill $\triangleright$ Update the assignment matrix
                    \ENDFOR

                    \STATE $\Lambda \gets (A^* \Lambda {A^*}^{\rm T}) \odot  (\mathbf{1} \mathbf{1}^{\rm T} - \mathbf{I}_{n \times n})$  $\triangleright$ \small{Update adjacency}
                    \STATE $I \gets A \ \hat{I}$  \hfill $\triangleright$  Update the reduce network current
                    \ELSE
                    \STATE $\Delta = 0$ \hfill $\triangleright$ If there is no feasible assignment
                    \ENDIF
            \ENDWHILE
    \end{algorithmic}
            \label{alg: exhaustive-search}
\end{algorithm}
We initialize the assignment set such that every node is a super-node and is assigned to itself. At each iteration, the set of candidate assignments must satisfy phase availability and connectivity constraints. The algorithm then evaluates the error of every potential assignment, selects the candidate with the minimum sum of MICE (SMICE) across all loading scenarios, and proceeds with updated adjacency, current vector, and assignment set.
For all iterations, $Y$ remains fixed, while the current vector $I^c$ is updated for each assignment. Solving $I^c = Y V^c$ results in the same super-node voltages that Kron reduction would yield.  
After each solve, reduced nodes are assigned the voltage of their super-node, both for new and previously reduced nodes.
The parallelizable part of the algorithm is the evaluation of candidate assignments, where each candidate assignment is evaluated independently with no data dependencies. This independence makes this block suitable for parallel computation of the resulting error of each candidate assignment (${\rm SMICE}$), which, in turn, makes the proposed approach scalable.
However, the Kron-based reduction of a radial network generates a smaller but dense meshed network~\cite{kron_dorfler,optikron2025}. To recover a radial three-phase distribution network, we now review and extend the radialization approach introduced in~\cite{optikron2025}.

\section{Three-phase Radialization}\label{sec:radialization}
Kron-based reduced networks are prone to being meshed and densely connected. 
This structure prevents the use of power flow algorithms optimized for three-phase radial networks. The density of the reduced networks increases the computational burden of subsequent optimization problems.
In \cite{optikron2025}, radialization was presented to find a radial equivalent of a single-phase Kron reduced network.
Here, we extend that approach to three-phase networks. 

\begin{figure}
    \centering
    \includegraphics[width=0.9\linewidth]{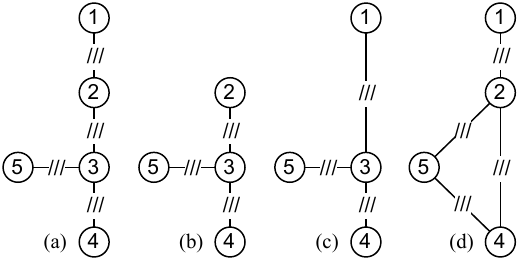}
    \caption{Change of topology with Kron reduction. While (a) shows the full network, (b), (c), and (d) represent the reduced networks, when nodes 1, 2, and 3 are Kron reduced, respectively. Kron reduction of node 3 with degree three results in a meshed network.}
    \label{fig: topology}
\end{figure}

The topology of a Kron reduced graph can be realized through its admittance matrix from~\eqref{eq: Y_kron}. In single-phase networks, any nonzero entry ${Y_{\rm Kron}}_{i,j}$ indicates a direct connection between nodes $i$ and $j$. For three-phase feeders, we extend this definition: a connection between nodes $i$ and $j$ exists if the corresponding $3\times3$ block in ${Y_{\rm Kron}}$ contains at least one nonzero entry. In the following, we first demonstrate that the topology of a three-phase Kron reduced network can be found through its Kron reduced graph Laplacian as Lemma~\ref{lemma: radial}. Then, we leverage this observation to extend single-phase radialization to three-phase cases. 
\begin{lemma} \label{lemma: radial}
    When Kron reduction is applied to a three-phase network and to a single-phase network with the same topology, the resulting reduced networks also yield the same topology.
\end{lemma}
\begin{proof}
Let $r\in \mathcal{V}$ be a three-phase node that is Kron reduced from a three-phase network with $Y \in \mathbb{C}^{3n\times 3n}$. 
For the sake of simplicity, assume all the nodes and lines in $Y$ are three-phase. We can partition the set of kept nodes $\mathcal{K} = \mathcal{V} \setminus r$ into $\mathcal{K}_a = \left\{ i \in \mathcal{K} \mid \Lambda_{i,r} = 1 \right\}$ and $\mathcal{K}_f = \left\{ i \in \mathcal{K} \mid \Lambda_{i,r} = 0 \right\}$. Now, we can further expand~\eqref{eq: Y-partitioned} to
\begin{align}
  Y
  \;=\;
  \begin{bmatrix}
    Y_{\mathcal{K}_{\text{f}},\mathcal{K}_{\text{f}}} 
      & Y_{\mathcal{K}_{\text{f}},\mathcal{K}_{\text{a}}} 
      & \mathbf{0} \\
    Y_{\mathcal{K}_{\text{a}},\mathcal{K}_{\text{f}}}  
      & Y_{\mathcal{K}_{\text{a}},\mathcal{K}_{\text{a}}}
      & Y_{\mathcal{K}_{\text{a}},r} \\
    \mathbf{0} 
      & Y_{r,\mathcal{K}_{\text{a}}} 
      & Y_{r, r}
  \end{bmatrix},
\end{align}
and~\eqref{eq: Y_kron} into
\begin{align} \label{eq: Kron_adj_far}
  Y_{\text{Kron}} 
  =
  \begin{bmatrix}
    Y_{\mathcal{K}_{\text{f}},\mathcal{K}_{\text{f}}}
      & Y_{\mathcal{K}_{\text{f}},\mathcal{K}_{\text{a}}}\\
    Y_{\mathcal{K}_{\text{a}},\mathcal{K}_{\text{f}}} 
      & Y_{\mathcal{K}_{\text{a}},\mathcal{K}_{\text{a}}} 
      - Y_{\mathcal{K}_{\text{a}},r} \ Y_{r,r}^+ \ Y_{r,\mathcal{K}_{\text{a}}}
  \end{bmatrix}.
\end{align}
From~\eqref{eq: Kron_adj_far}, we can infer that the only part of the reduced network that may have different topology from the full network is the connection among nodes in $\mathcal{K}_a$. 
Since the original admittance matrix $Y$ corresponds to a radial network and no two distinct neighbors of $r$ are directly connected, $Y_{\mathcal{K}_a,\mathcal{K}_a}$ is block diagonal of size $3|\mathcal{K}_a|\times 3|\mathcal{K}_a|$.  On the other hand, every node in $\mathcal{K}_a$ is adjacent to $r$, which implies that all the elements in $Y_{\mathcal{K}_a,r} \in \mathbb{C}^{3|\mathcal{K}_a|\times 3}$, $Y_{r,\mathcal{K}_a} \in \mathbb{C}^{3 \times 3|\mathcal{K}_a|}$, and $Y_{r,r}^+ \in \mathbb{C}^{3 \times 3}$ have non-zero values. Therefore, $Y_{\mathcal{K}_a,r} \, Y_{r,r}^+ \, Y_{r,\mathcal{K}_a}$ and, in turn, the lower right block of~\eqref{eq: Kron_adj_far} yields a fully dense sub-matrix with all the values being non-zeros.
This implies that, after eliminating node $r$, all nodes in $\mathcal{K}_a$ become mutually connected, while the connections among nodes in $\mathcal{K}_f$, remain unchanged.
The same structural effect occurs in the single-phase Kron reduction~\cite{optikron2025}. Hence, Kron reduction of a three-phase network and a single-phase network with the same topology, e.g., its graph Laplacian, results in reduced networks with identical topologies.
\end{proof}
Fig.~\ref{fig: topology} demonstrates different stages of Kron reduction.
Reference~\cite{inverse_pf} proves that Kron reduction of radial networks is an edge-disjoint union of maximal cliques. 
Lemma~\ref{lemma: radial} extends this property to three-phase networks. A clique is a sub-graph that all the nodes in the sub-graph are mutually connected. When a clique is not included in any larger clique, it is called a maximal clique. Any maximal cliques with three or more nodes creates meshes in Kron reduced networks, which is the consequence of eliminating nodes with degree $\ge 3$. These properties are identical between single and three-phase networks. Therefore, we summarize the radialization procedure of~\cite{optikron2025}, which can be extended directly into three-phase networks. 

Radialization consists of identifying and re-inserting critical nodes for each maximal clique in a Kron reduced network. Critical nodes are defined as those with degree $\ge 3$ in a sub-tree of the original feeder that spans the nodes of each maximal clique. Adding these nodes to the set of super-nodes ensures that the reduced network is radial. 
This does not mean all nodes with degree $\ge 3$ must be reinserted, only the minimal subset whose absence would otherwise create cliques. Thus, if $\mathcal{K}$ is the set of super-nodes in the reduced meshed network and $\mathcal{R}_c$ is the set of critical nodes for all the maximal cliques, then the set of super-nodes in the radialized network is $\mathcal{K}_{\rm radial} = \mathcal{K} \cup \mathcal{R}_c$.
Finally, if injections at critical nodes are not reassigned, the pre-radialized and radialized networks are electrically equivalent in terms of the voltage profile at the super-nodes.

\section{Simulation and Results}\label{sec:results}
To evaluate the performance of the proposed framework, simulations are carried out on two realistic three-phase unbalanced feeders from Vermont, USA: feeder A with 8,381 nodes and feeder B with 5,591 nodes. To investigate performance across different network scales, multiple sub-feeders of varying sizes (100, 300, 500, and 1000 nodes) are extracted from these systems. For each network, two representative operating scenarios are considered. The two scenarios correspond to the highest and lowest loading conditions observed over a week of historical hourly data, i.e., 168 loading profiles. All computations were performed on the Vermont Advanced Computing Center (VACC) at the University of Vermont. The simulation codes used in this study were implemented in Julia 1.11.5. To solve the optimization problems, JuMP v1.29.1~\cite{JuMP} and the Gurobi Optimizer 12.0 were employed, with the optimality gap set to 0\%. The GPU implementation was carried out using the CUDA package~\cite{cuda}.

\subsection{Comparison: MILP vs. exhaustive search}
This subsection compares the performance of the MILP and exhaustive-search formulations.
Table~\ref{tab:q_effect} summarizes the effect of the reduction limit $q$ on the MILP-based reduction applied to the 100-node sub-feeders of A and B for $\bar{E}=0.001$.
\begin{table}[!t]
\centering
\caption{Effect of the reduction limit $q$ on MILP performance for $\bar{E}=0.001$ (100-node sub-feeders).}
\begin{tabular}{c c c c}
\toprule
$q$ & Reduction (\%) & Max $|\Delta V|$ (p.u.) & SMICE$_\text{abs}$ \\
\midrule
\multicolumn{4}{l}{{Sub-feeder A}}\\
1   & 70.0 & 0.0010 & 0.0221 \\
3   & 70.0 & 0.0009 & 0.0179 \\
10  & 71.0 & 0.0010 & 0.0276 \\
20  & 70.0 & 0.0010 & 0.0220 \\
\midrule
\multicolumn{4}{l}{{Sub-feeder B}}\\
1   & 68.0 & 0.0010 & 0.0142 \\
3   & 68.0 & 0.0010 & 0.0138 \\
10  & 68.0 & 0.0010 & 0.0146 \\
20  & 69.0 & 0.0010 & 0.0191 \\
\bottomrule
\end{tabular}
\label{tab:q_effect}
\end{table}
Across each test feeder, the reduction levels and voltage errors ($ 0.001$ p.u.) are almost identical, which emphasizes that increasing 
$q$ provides no meaningful gain in accuracy or reduction capability.
On the other hand, small variations in SMICE suggest that larger $q$ values may slightly increase intra-cluster error due to reduced exploration of feasible assignments in each iteration. The remainder of the analysis therefore focuses exclusively on $q=1$, which provides minimal computational burden.

\begin{figure}
    \centering
    \includegraphics[width=1\linewidth]{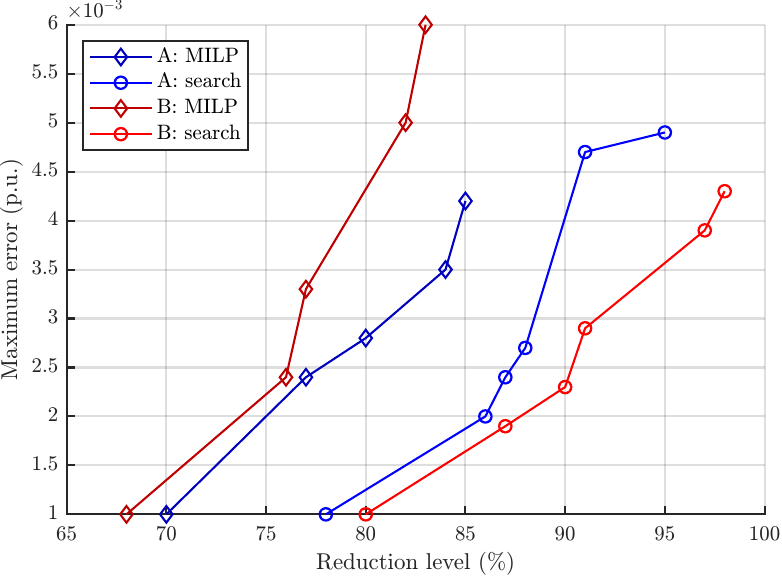}
    \caption{Comparison of reductions based on MILP and exhaustive search on 100-node sub-feeders: maximum voltage magnitude error versus reduction.}
    \label{fig: 100node_comparison}
\end{figure}

Figure~\ref{fig: 100node_comparison} compares the MILP-based and exhaustive search formulations on the 100-node sub-feeders.
The MILP formulation relies on decomposition into rectangular coordinates for complex variables, which degrades the accuracy. The exhaustive search, by operating directly on complex values, achieves lower voltage errors and allows higher reduction levels.
In early iterations, the error surface is typically flat, meaning many candidate reductions yield negligible voltage error. For example, in the first iteration of Opti-KRON for network A, there are 32 distinct reductions whose effects on the objective function are nearly indistinguishable ($<10^{-6}$). This flatness allows the optimizer to select among multiple feasible reductions with similar accuracy, which alters the adjacency matrix across iterations and can lead to different reductions. The MILP formulation serves as a methodological baseline. However, its combinatorial complexity limits scalability to large three-phase feeders. For larger feeders, we therefore rely on the exhaustive-search approach, which scales significantly better in practice.

\subsection{Accuracy and generalization}
\begin{figure}
    \centering
    \includegraphics[width=1\linewidth]{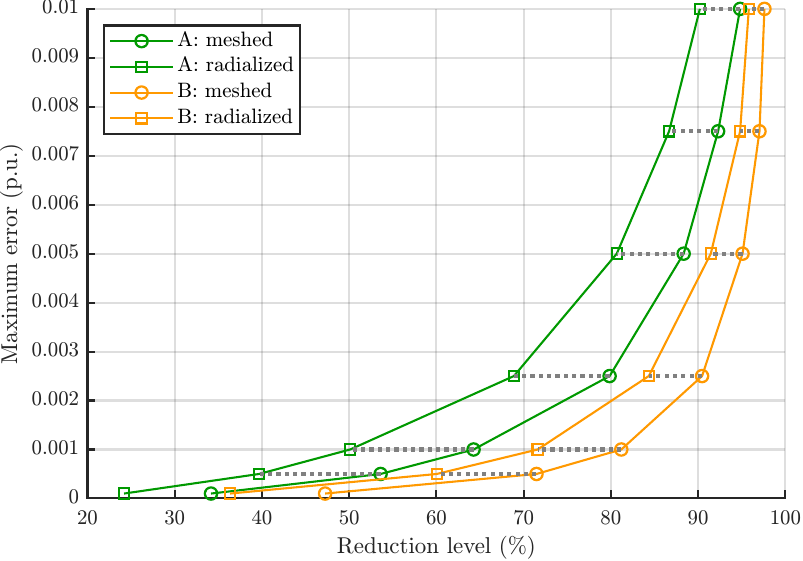}
    \caption{Reduction ratio versus maximum voltage deviation for the full A and B feeders. Radialization of highly reduced networks decreases reduction level by $2 - 3\%$, while for moderately reduced networks the change is $10 - 12\%$. Radialized and pre-radialized networks have identical accuracy.}
    \label{fig: full_errorr}
\end{figure}

In this subsection, the accuracy and robustness of the reduced networks of the full Vermont feeders are investigated. 
Figure~\ref{fig: full_errorr} illustrates the trade-off between network reduction and voltage accuracy for the full A and B feeders. 
For meshed reduced networks, each point represents the converged reduced network from the exhaustive search for a voltage-deviation limit $\bar{E}\in[0.0001,0.01]$~p.u. For example, $\bar{E}=0.001$~p.u. yields 65\% and 81\% reduction for networks A and B, respectively.
Both networks demonstrate a monotonic decrease in accuracy with increasing levels of reduction.

The difference in reduction between the two networks for the same error threshold arises from the distinct network structures and loading conditions. These differences lead to variations in nodal voltages and different sensitivities to reduction.

\begin{figure}
    \centering
    \includegraphics[width=1\linewidth]{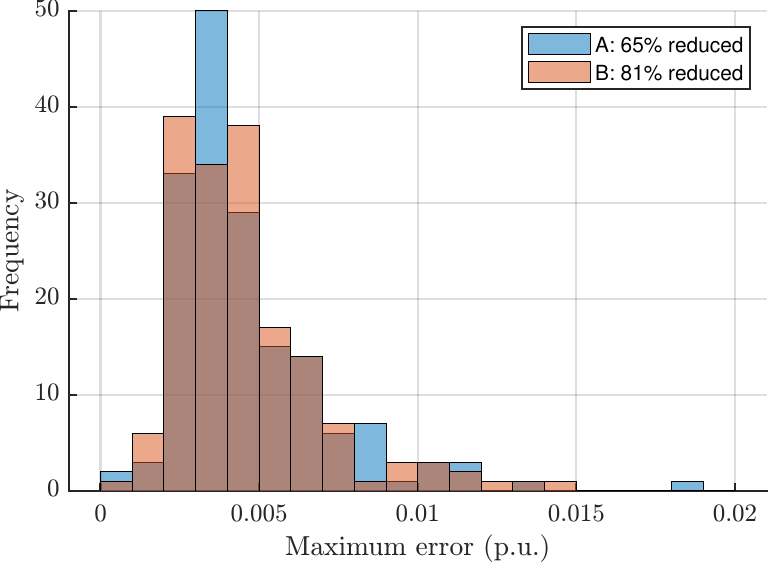}
    \caption{Distribution of maximum voltage errors across hourly loading scenarios for the reduced A and B networks.}
    \label{fig: robustness}
\end{figure}

To assess the robustness of the reduced models to varying operating conditions, we evaluate their performance across a set of historical hourly loading scenarios over one week. 
Although the reduction process is performed using only two loading profiles, the resulting reduced networks remain accurate for many operating points. Figure~\ref{fig: robustness} represents the histogram of maximum error across different loading scenarios. 

\subsection{Computational performance and scalability}
This subsection analyzes the computational efficiency and scalability of the exhaustive search on CPU and GPU.
To find the reduced networks of the full feeders, the exhaustive-search algorithm was executed on GPU, and the corresponding performance results are reported in Table~\ref{tab: sim_time_reduction}.
\begin{table}[t]
\centering
\caption{Simulation time vs.\ reduction level for A and B networks using exhaustive search on GPU.}
\label{tab: sim_time_reduction}
\begin{tabular}{@{}c c c@{}}
\toprule
\multirow{2}{*}{Reduction (\%)} & \multicolumn{2}{c}{\textbf{Computation time (s)}} \\
\cmidrule(lr){2-3}
 & A (8381 nodes) & B (5991 nodes) \\
\midrule
10   & 562.6 & 240.2 \\
20   & 1075.5 & 459.7 \\
30   & 1540.1 & 663.7 \\
40   & 1968.1 & 885.4 \\
50   & 2401.2 & 1034.3 \\
60   & 2755.9 & 1200.5 \\
70   & 3113.2 & 1355.1 \\
80   & 3400.4 & 1497.6 \\
90   & 3551.0 & 1631.7 \\
\bottomrule
\end{tabular}
\end{table}

To compare the performance of the model on GPU and CPU, Table \ref{tab: simtime} reports the simulation time for different network sizes (100, 300, 500, and 1000 nodes) and reduction levels (25, 50, and 75\%).
\begin{table}[t]
\centering
\caption{Simulation time for exhaustive reduction algorithm~\ref{alg: exhaustive-search} for different networks and reduction levels}
\label{tab: simtime}
\begin{tabular}{cc|cccc}
\toprule
\multirow{2}{*}{Device} & \multirow{2}{*}{Reduction} & 
\multicolumn{4}{c}{\textbf{Computation time (s)}} \\
\cmidrule(lr){3-6}
 &  & 100-n & 300-n & 500-n & 1000-n \\
\midrule
\multirow{3}{*}{CPU} 
& 25\% & \textbf{0.2}  & 6.5  & 54.3  & 744.5 \\
& 50\% & \textbf{0.4}  & 10.5 & 94.2  & 1379.6 \\
& 75\% & \textbf{0.5}  & 12.4 & 124.4 & 1850.2 \\
\midrule
\multirow{3}{*}{GPU} 
& 25\% & 0.8  & \textbf{4.77} & \textbf{10.94} & \textbf{53.7} \\
& 50\% & 1.3  & \textbf{7.21} & \textbf{17.46} & \textbf{84.3} \\
& 75\% & 1.6  & \textbf{9.4}  & \textbf{23.4}  & \textbf{106.6} \\
\midrule
& \textbf{Avg. GPU speedup}  & 0.3X & 1.4X  & 5.2X  & 15.9X  \\
\bottomrule
\end{tabular}
\end{table}
During the initial iterations, both implementations exhibit higher computation time because identifying the next feasible assignment requires exploring a larger search space. 
For smaller networks, the CPU completes the reduction faster due to lower parallelization overhead. 
For larger networks (with at least 300 nodes), the GPU achieves substantial acceleration and exhibits scalability with network size, whereas the CPU runtime grows rapidly and becomes impractical for large feeders.

\section{Conclusion and Future Work}\label{sec:end}
This work extended the network-reduction framework previously developed for single-phase feeders to unbalanced three-phase distribution systems. The proposed formulation enables accurate aggregation of large feeders while maintaining the voltage profile of the original networks across all phases. By adopting a direct exhaustive-search approach and leveraging parallel execution on GPUs, the method achieves high reduction levels with negligible voltage deviation. We applied the proposed framework to two real-world unbalanced three-phase networks and demonstrated the scalability and practicality of the technique.

Future work will generalize the Opti-KRON framework to transmission systems, with emphasis on preserving line flow accuracy between the reduced and original networks. We are also interested in investigating how optimization and control tasks---such as optimal power flow and hosting capacity assessment---performed on reduced models can be consistently lifted back to the full network. 

\printbibliography
\end{document}